\documentstyle[preprint,aps]{revtex}

\newcommand{\cs}{C$_{60} $}
\newcommand{\csm}{C$_{60}^{-} $}

\newcommand{\ba}{\begin{array}}
\newcommand{\ea}{\end{array}}
\newcommand{\beq}{\begin{equation}}
\newcommand{\beqa}{\begin{eqnarray}}
\newcommand{\eeq}[1]{\label{#1}\end{equation}}
\newcommand{\eeqa}[1]{\label{#1}\end{eqnarray}}

\newcommand{\refp}[1]{(\ref{#1})}

\newcommand{\ket}[1]{\left| \left. #1 \right \rangle \right.}

\newcommand{\ev}[3]{\left \langle #1\left |#2\right |#3\right \rangle}

\begin{document}
\title{Electron--Phonon Coupling in Charged Buckminsterfullerene}

\author{N. Breda$^{1,2}$, R.A. Broglia$^{1,2,3}$, G. Col\`o$^{1,2}$, H.E. 
Roman$^{1,4}$, F. Alasia$^{1,2}$,\\ G. Onida$^5$, V. Ponomarev$^6$ and E. Vigezzi$^2$ }

\address{$^1$Dipartimento di Fisica, Universit\`a di Milano,
Via Celoria 16, I-20133 Milano, Italy.}

\address{$^2$INFN, Sezione di Milano, Via Celoria 16, I-20133 Milano, Italy.}

\address{$^3$The Niels Bohr Institute, University of Copenhagen,
2100 Copenhagen, Denmark.}

\address{$^4$Institut f\"ur Theoretische Physik III, Universit\"at Giessen,
Heinrich-Buff-Ring 16, \\ 35392 Giessen, Germany.}

\address{$^5$Dipartimento di Fisica, Universit\`a di Roma Tor Vergata,
Via della Ricerca Scentifica 1,\\ I-00133 Roma, Italy.}

\address{$^6$Laboratory of Theoretical Physics, Joint Institute for Nuclear
Research, Dubna, \\
Head Post Office, P. O. Box 79, Moscow, Russia.}
\date{\today}

\maketitle

\smallskip


\begin{abstract}
A simple, yet accurate solution of the electron--phonon coupling problem in \cs \ is presented. 
The basic idea behind it is to be found in the parametrization of the ground
state electronic density of the system calculated making use of ab--initio methods, in term of 
sp$^{2+x}$ hybridized orbitals. This parametrization allows for an economic 
determination of the deformation potential associated with the fullerene's normal modes. The 
resulting electron--phonon coupling constants are used to calculate Jahn--Teller effects 
in \csm, and multiple satellite peaks in the corresponding photoemission reaction. Theory 
provides an accurate account of the experimental findings.
\end{abstract}

\newpage
\narrowtext
The discovery of fullerenes \cite{refe_1}, and the subsequent ``mass
synthesis'' \cite{refe_2} of these hollow molecules, have prompted the
search for a new class of materials such as fullerides (salts like 
K$_3$C$_{60}$), fullerites (\cs\  molecular crystals), etc., in which
fullerenes play the role of the building blocks. The understanding of the
response of these building blocks to external fields is essential in the
characterization of the associated new materials. A central issue in this
quest is how the electronic 
properties of fullerenes are modified by the coupling of 
electrons to phonons. This question can be answered in terms of ab--initio methods 
(cf., e.g., Refs. \cite{terza,quarta} and Refs. therein; cf. also Ref. \cite{Gel94}). 
These methods are, however, computationally demanding and not particulary
transparent, and much
of the physical insight is lost in the complexity of
mathematics. 

In the present paper we present a simple, yet accurate solution of the electron--phonon 
coupling problem particularly suited for fullerenes, and apply it to C$_{60}^{-}$. 
The central  
idea behind this solution is based on a parametrization of the ground
state electronic density of the system calculated making use of ab--initio methods in terms of 
hybridized 
atomic orbitals. In particular, in the case of C$_{60}$ 
fullerene, of sp$^{2+x}$ orbitals.  
Such parametrization allows for an economic determination of the deformation potential 
associated with the different normal modes. The resulting electron--phonon coupling constants 
are used to calculate Jahn--Teller effects in C$_{60}^{-}$ and multiple satellite peaks in 
the corresponding photoemission spectrum. The resulting cross section agrees well with the 
experimental data, better than  
any of the cross sections obtained making use of the 
electron--phonon coupling constants available in the 
literature~\cite{terza,quarta,cita,citb,citc}. 
Furthermore, the 
extension of the hybrid orbital model to describe the electron--phonon 
coupling phenomenon in fullerenes other than C$_{60}$ as well as in nanotubes 
is simple to carry out.

The electron--phonon coupling is determined by the matrix element of the deformation potential
\beq
\left. V_{\mbox{\tiny{def}}}(\vec r)  =  \sum_{\beta}\sum_{k=1}^{3}
\left( Q_{\beta} \right)_{k} {\left( \nabla_{\beta} \right)}_{k} 
V_{\mbox{\tiny{e}}} (\vec r, \{ R \})
\right|_{\{ R\} = \{R^0\}},
\eeq{uno}
where $\{R\}$ represents the set of ionic coordinates with equilibrium values $\{R^0\}$, 
$\left( Q_{\beta} \right)_{k}$ the displacement field associated with the normal modes of the 
$\beta$-ion in the $k$=($x$,$y$,$z$) directions, while $V_{\mbox\tiny{e}}$ is the
total potential acting on the electrons. This potential can be expressed, in the local density 
approximation (LDA), as
\beqa
V_{\mbox{\tiny{e}}}(\vec r,\{R\}) & = & 
V_{\mbox{\tiny{pseudo}}}^{L}(\vec r,\{R\})+
V_{\mbox{\tiny{Hartree}}}(\rho(\vec r,\{R\}))+ \nonumber \\ 
& & V_{\mbox{\tiny{xc}}}(\rho(\vec r,\{R\}))+
V_{\mbox{\tiny{pseudo}}}^{NL}(\vec r,\{R\}).
\eeqa{aggiunta}
The first three terms are local and arise from: a) the pseudopotential associated with the
ions; b) the Coulomb interaction acting among the electrons (Hartree field); 
c) exchange (Fock field) and correlation effects. The last term in Eq. 
\refp{aggiunta} reflects the non--local part of the ionic pseudopotential. 

Because the first and the last terms of Eq. \refp{aggiunta} display simply
an explicit dependence on the ion 
positions, the calculation of their gradient (cf. Eq. \refp{uno}) presents no 
difficulties. This is not the case for the second and third terms of Eq. \refp{aggiunta}, 
which depend on $\{R\}$ implicitly via the electronic density, a fact which 
can be used to write
\beq
\sum_{\beta} \vec Q_\beta \cdot {\vec \nabla_\beta} V_{i}[\rho(\vec r,\{R\})] = 
\sum_{\beta} \frac{\partial V_{i}[\rho]}{\partial \rho} \vec Q_\beta \cdot 
{\vec \nabla_\beta} \rho(\vec r,\{R\}),
\eeq{dercomp}
where $i$=(Hartree,xc). Because in the LDA there is an explicit relation between $V_i$ 
and $\rho$, the derivative $\partial V_{i}/\partial \rho$ can be calculated analytically. 
Consequently, the basic difficulty associated with the calculation of the deformation 
potential defined in Eq. \refp{uno}, and thus of the electron--phonon coupling constants, 
lies in  the calculation of the gradient of the electronic density along the normal 
displacements. To overcome this difficulty, we shall parametrize the LDA results in term 
of sp$^{2+x}$ hybrid orbitals, in such a way that three of the four orbitals are 
directed along the carbon bonds of fullerene \cs, while the fourth takes care 
of the $\pi$-bonding present in the hexagons and 
is directed essentially perpendicularly to the fullerene surface.

In keeping with the fact that in \cs\  each atom has three nearest 
neighbours, the hybrid orbitals we are interested in can be written as

\beqa
|\phi_1> & = & a_1|s>+b_1|p_x>+c_1|p_y>+d_1|p_z>,  \nonumber \\
|\phi_2> & = & a_2|s>+\delta_2(m_2|p_x>+n_2|p_y>+t_2|p_z>), \label{ibridi} \\ 
|\phi_3> & = & a_3|s>+\delta_3(m_3|p_x>+n_3|p_y>+t_3|p_z>), \nonumber \\
|\phi_4> & = & a_4|s>+\delta_4(m_4|p_x>+n_4|p_y>+t_4|p_z>). \nonumber
\end{eqnarray}
Here $m_{j}=\cos \alpha_{j}$, $n_{j}=\cos \beta_{j}$, 
$t_{j}=\cos \gamma_{j}$, and $\alpha_{j}$, $\beta_{j}$, $\gamma_{j}$ ($j$=2,3,4) are the angles 
which define the direction of the bond between a Carbon atom and each of its three nearest 
neighbours, in a system of reference centered on the atom. After having fixed the direction 
of these orbitals, there still remain ten free parameters in Eq. \refp{ibridi}, parameters 
which are completely determined by the orthonormalization condition. 

To describe  the radial dependence of the $\ket{s}$ and $\ket{p}$ orbitals, we have used 
the functions $R_{s}  =  \frac{2}{\sqrt{\sigma_{1}^{3}}}e^{(-r/\sigma_{1})}$ and 
$R_{p}  =  \frac{2}{\sqrt{3\sigma_{2}^{5}}}re^{(-r/\sigma_{2})}$, usually 
employed in the description of the Carbon atom. We have however adjusted the parameters $\sigma_{1}$ and 
$\sigma_{2}$ in order to obtain the best fit to the LDA \cs-density (see, for
instance, Ref.\ \cite{Ala94}). One can then write the contribution to 
the total density arising from a single atom and, with the help of standard techniques, 
carry out a multipole expansion of this contribution around the center of the molecule. 
Adding the contributions of the 240 electrons one obtains the total density.

In Fig.~\ref{fig1} we display the two lowest multipoles of the \cs \
ground state density, calculated in the LDA including exchange--correlation effects 
according to the parametrization of Perdew and Zunger \cite{perzu}. The role of the 
Carbon atoms were taken into account in terms of norm--conserving 
pseudopotentials \cite{perpse}. In the same figure we show the results of the hybrid orbital 
model, for $\sigma_{1}=0.78\ \AA$ and $\sigma_{2}=0.31\ \AA$ \cite{testo}. 

The next step consisits in the calculation of the gradient of the density and, through 
Eqs. \refp{dercomp} and \refp{ibridi}, the deformation potential. Within the hybrid 
orbital model, moving around the ions change the direction of the orbitals but not 
their shape. This means that the weights $\delta_{j}$ in Eq. \refp{ibridi} are 
fixed, and the only quantities which change are the angles $\alpha_{j}$, 
$\beta_j$ and $\gamma_j$. The calculation of ${\vec \nabla}\rho(\vec r,\{R\})$
becomes then quite simple.

In Fig. \ref{fig2} we display the two lowest, and most important, multipole contributions 
of the local part of the deformation potential of \cs \ for 
the lowest A$_g$ phonon, 
corresponding essentially to a breathing mode of the system, as calculated in the LDA. 
The wavenumber of this mode is equal to 491 cm$^{-1}$ and to 496 cm$^{-1}$
for \cs\  in solution and in the solid phase, respectively
\cite{exper1,exper2}. The wavenumber and zero point motion of the isolated
molecule, calculated in the bond--charge
model, are 493 cm$^{-1}$ and $53.4\;10^{-3} \AA$, respectively \cite{Oni92}. 
As seen from Fig.~\ref{fig2}, 
the hybrid orbital model provides an overall account of the ab--initio results. 
The discrepancies observed for small values of $r$ for the multipole (L,M)=(0,0)
have little influence on the corresponding matrix elements, because the electronic 
wavefunctions are quite small around the origin of the molecule.

Making use of these results, one can calculate the electron-phonon coupling 
matrix elements in C$_{60}$. These are the matrix elements needed, for example, 
in the evaluation of the lineshape of allowed as well as of forbidden 
electronic transitions in C$_{60}$. To carry out similar calculations in 
C$_{60}^-$ one should employ a deformation potential which is evaluated by 
making use of the electronic density of the negative ion, at the ground state 
geometry. Because the density of the 240 valence electrons of C$_{60}$ is 
not appreciably altered by adding one more electron, one expects the 
deformation potentials associated with C$_{60}$ and C$_{60}^-$ to be quite 
similar. In fact, we have carried out fully relaxed, ab-initio calculations 
of the matrix elements in C$_{60}^-$ and found that they agree with those 
of C$_{60}$ within less than 10\%. In keeping with these results, 
the electron-phonon matrix elements calculated 
starting from the C$_{60}$ elecronic density and ground state ionic
configuration are used in the following.    

In \csm \ the state t$_{1u}$ is occupied with a single electron. This level is 
separated by an energy of the order of the eV from neighbouring levels, while the 
electron--phonon coupling matrix elements 
to be found below are of the order of the meV. Consequently, 
it seems justified to consider, within the present context, that the electronic motion 
is confined to the t$_{1u}$ level. Under such circumstances, and because of symmetry 
reasons, the only possible couplings are to phonons with A$_g$ and H$_g$
symmetries (cf., e.g., Ref.\ \cite{citb}).  

The matrix elements $\ev{t_{1u} \nu}{V_{\mbox{\tiny{def}}}}{t_{1u}}$, where $\nu$ 
stands for the quantum numbers of the phonon, are related to the partial 
electron--phonon coupling constant $\lambda_{\nu}/{N(0)} = \alpha {g_{\nu}^{2}}/{\omega_{\nu}}$, 
according to $\ev{t_{1u}
\nu}{V_{\mbox{\tiny{def}}}}{t_{1u}}=(g_{\nu}/2)\ W_{nm}^{l}$ (cf., e.g., 
Ref.\ \cite{Gunn2} and Refs. therein). In the above expression 
N(0) is the density of levels at the Fermi energy, $\alpha$ is equal to 1/3 for A$_g$ 
phonons and 5/3 for H$_g$ phonons, while $\omega_{\nu}$ is the energy of the phonon. 
The quantities $W_{nm}^{l}$ are geometric coefficients, the index $l$ 
distinguishing between the 
different degenerate states of each phonon (H$_g$ is five--fold degenerate while 
A$_g$ is single--fold degenerate). 

In Table \ref{tab1} we display the multipole contributions to the matrix element 
$\ev{t_{1u} A_{g}}{V_{\mbox{\tiny{def}}}}{t_{1u}}$, associated with the lowest energy 
A$_g$ mode calculated making use of the LDA and of the hybrid orbital 
model. The different contributions of the model display  the same sign and similar 
order of magnitude as those calculated in LDA, while the summed contribution agree 
within 20$\%$.

Following the same steps as those leading to the results displayed in table \ref{tab1}, 
the different matrix elements $\ev{t_{1u} \nu}{V_{\mbox{\tiny{def}}}}{t_{1u}}$ \ 
($\nu$=A$_g$,H$_g$) have been calculated. Our results have been compared with
those  
from other theoretical calculation available in the 
literature~\cite{terza,quarta,citb,citc}, as well as with 
the empirical values obtained from Gunnarsson's systematic analysis of the 
photoemission spectra of \csm \ \cite{cita}. While this analysis  
indicates that the coupling of the t$_{1u}$ electron to the H$_g$(2) leads to the largest value of 
$\lambda_{\nu}/N(0)$, the results reported in 
Refs.~\cite{terza,quarta,citb,citc} indicate the coupling to the 
H$_g$(7)--phonon to be the most important.
In the hybrid orbital model discussed above, 
the largest coupling of the t$_{1u}$ level is to the H$_g$(2)--phonon, 
in agreement with the analysis by Gunnarsson and co-workers \cite{diciass}. 

Making use of the matrix elements $\ev{t_{1u} \nu}{V_{\mbox{\tiny{def}}}}
{t_{1u}}$ ($\nu$=A$_g$,H$_g$) calculated in the hybrid orbital model, and of the results 
of the bond charge model \cite{Oni92} to describe the properties of the
phonons, we have solved the total electron--phonon Hamiltonian containing an electronic term, 
a phonon term, and a linear coupling term (cf., e.g., Ref. \cite{cita}) in a basis of 
one t$_{1u}$ electron and up to three phonon states. 
The lowest eigenvalue $\widetilde{\ket{t_{1u}}}=C^{(0)}\ket{t_{1u}}+\sum_{\nu}C_{\nu}^{(1)} 
\ket{t_{1u} \otimes \nu} + \ldots$
was calculated using the Lanczos method. The first term in $\widetilde{\ket{t_{1u}}}$ 
describes a state with no phonons, the second term a state with one phonon, etc. Making 
use of these results, we have calculated the photoemission cross section assuming the 
sudden approximation, where the emitted electron does not interact with the system 
left behind \cite{cita}.

In Fig. \ref{fig3} we show the results of the hybrid orbital model, in comparison with the 
results of the analysis of the photoemission data carried out by Gunnarsson and
co-workers \cite{cita}. Although 
the hybrid orbital model leads to a somewhat weaker electron--phonon 
coupling than required by the experimental finding, and consequently to a somewhat too 
large value of $C^{(0)}$, it provides a much better account of the empirical spectrum 
than the other theoretical models, whose partial electron--phonon coupling constant
have been reported in
Refs.~\cite{terza,quarta,citb,citc} (cf. also table I of Ref. \cite{cita}).

We conclude that the hybrid orbital model of the electron--phonon coupling 
 displays a number of attractive features: 
i) it leads to matrix elements of the deformation potential which reproduce
quite accurately the results of ab--initio calculations; ii) it provides an excellent account 
of the photoemission spectra of \csm; iii) it is quite economic to use, and can be extended 
at profit to fullerenes other than \cs \ as well as to nanotubes.

Finantial support by NATO under grant CRG 940231 is gratefully acknowledged. 
We are also indebted for support to INFM Advanced Research Project
``CLASS''.

\narrowtext 
\begin{table}[h]
\caption{ Different multipole contributions to the matrix element 
$\ev{t_{1u} A_{g}}{V_{\mbox{\tiny{def}}}}{t_{1u}}$, associated with the lowest energy 
A$_g$ mode of \cs \ ($t_{1u}$ is the LUMO of the molecule). }

\vspace{0.3cm}

\begin{tabular}{ccc}
          & \multicolumn{2}{c}{Matrix Element [meV]} \\
 L=       &      LDA             & Hybrid Orbital Model  \\
 \tableline
 0        & -47.739              & -49.506 \\
 6        &  -3.319              &  -2.369 \\
 10       &  19.741              &  20.275 \\ 
 12       &  -1.030              &  -2.828 \\ 
 16       &   6.082              &   5.869 \\ 
 18       &  14.940              &  15.473 \\ 
 20       &   2.139              &   1.777 \\ 
Total     &  -9.399              & -11.309 \\
\end{tabular}

\label{tab1}
\end{table}
\narrowtext

\narrowtext


\begin{figure}
\caption{ Comparison between the results of LDA (full lines) and Hybrid
Orbital Model (dashed lines) for the two main multipole contributions to 
the ground state density of \cs. (a) and (b) refer to the 
(L,M)=(0,0) and (6,0) contributions respectively. }
\label{fig1}
\end{figure}

\begin{figure}
\caption{ Same as Fig.~\ref{fig1} for the two main multipole contributions 
to the deformation potential associated with the lowest A$_g$ phonon of 
\cs. } 
\label{fig2}
\end{figure}

\begin{figure}
\caption{ Results for the photoemission spectrum of \csm, obtained making 
use of the electron-phonon matrix elements calculated in the present paper as 
well as in previous theoretical 
works~{\protect \cite{terza,quarta,citb,citc}}. The
solid curve correspond to the experimental results~{\protect \cite{cita}}. } 
\label{fig3}
\end{figure}


\begin{thebibliography}{99}
\bibitem{testo}
These values can be compared to the atomic values, $\sigma_{1}=0.65 \AA$ and 
$\sigma_{2}=0.17 \AA$, and testify to the effect that nearest neighbour atoms 
have on the extension of the wavefunction of each C--atom in \cs.
\bibitem{diciass}
The electron--phonon coupling constant $\lambda_{\nu}/N(0)$ obtained by 
Gunnarsson from the analysis of the photoemission experimental data are, in
meV : 0.023 (H$_g$(8)), 0.017 (H$_g$(7)), 0.005 (H$_g$(6)), 0.012
(H$_g$(5)), 0.018 (H$_g$(4)), 0.013 (H$_g$(3)), 0.040 (H$_g$(2)), 0.019
(H$_g$(1)), 0.011 (A$_g$(2)), 0.000 (A$_g$(1)). The corresponding result of
the hybrid orbital model are, in the same units: 0.008, 0.016, 0.003, 0.006,
0.005, 0.011, 0.026, 0.010, 0.006, 0.001 respectively.
\end{thebibliography}
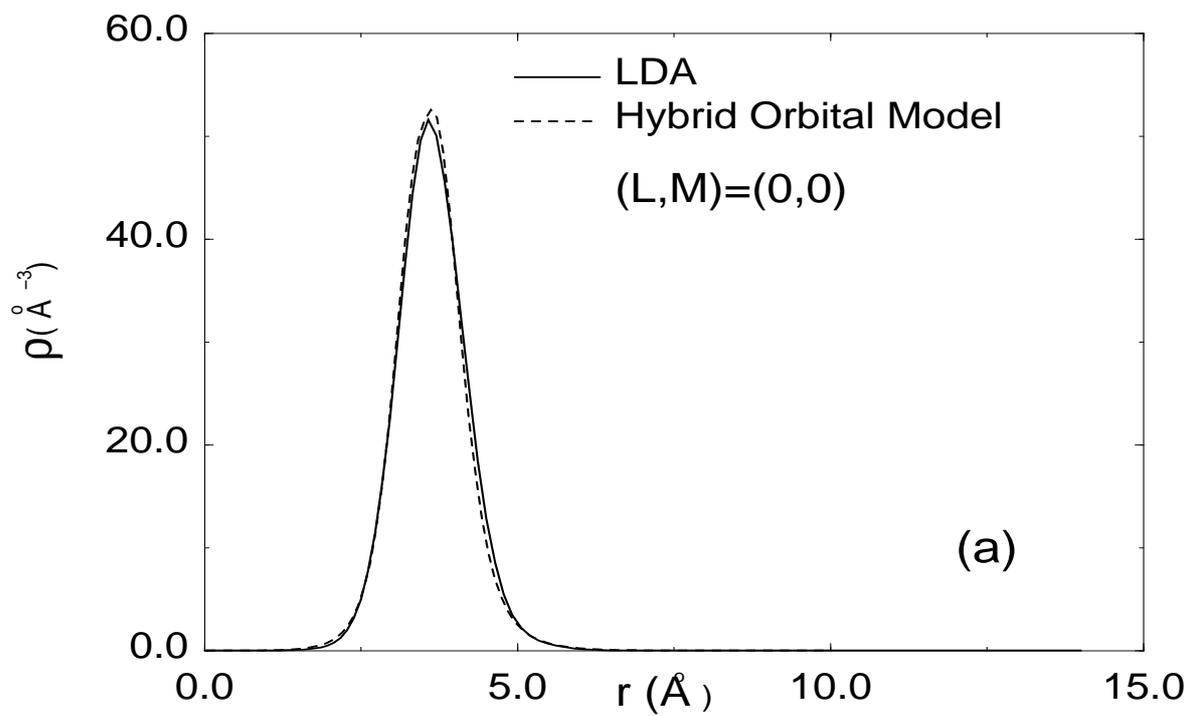
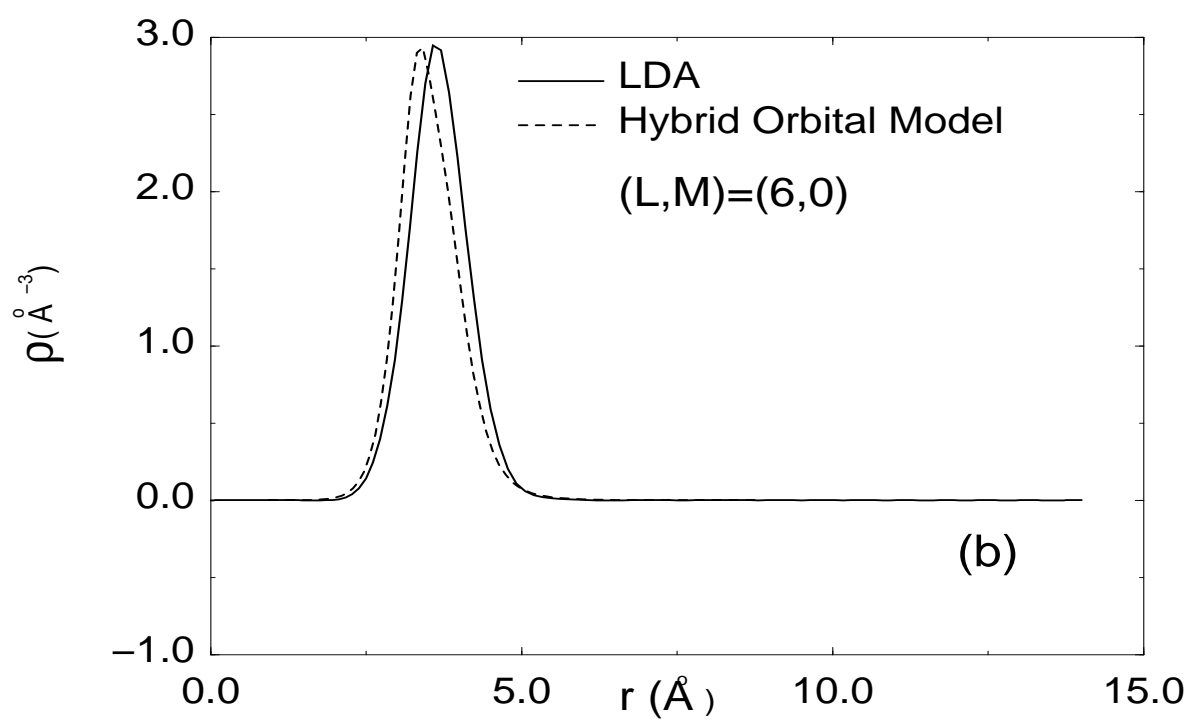
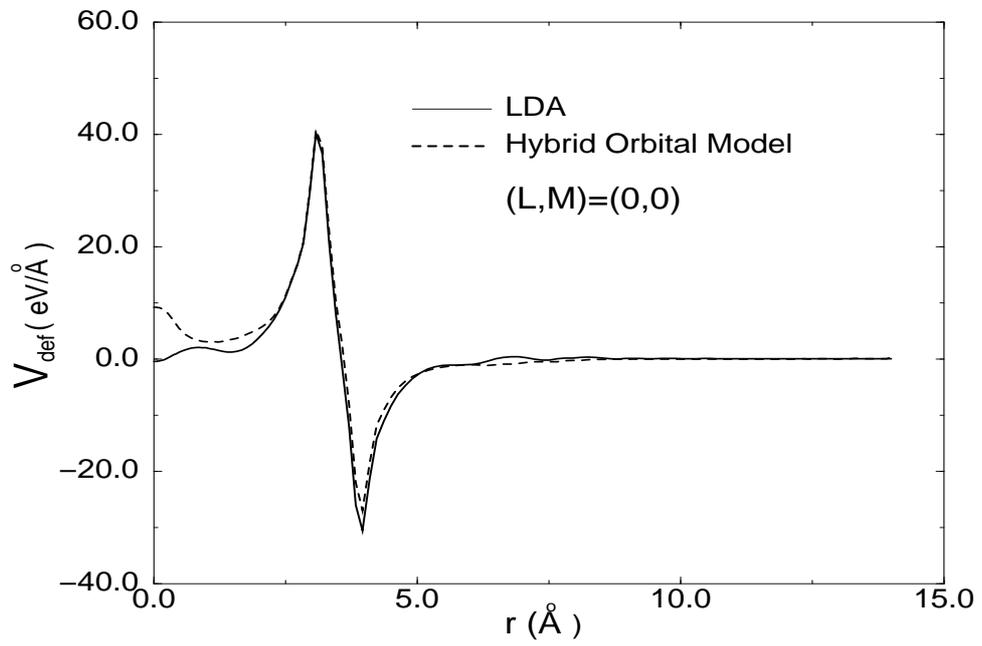
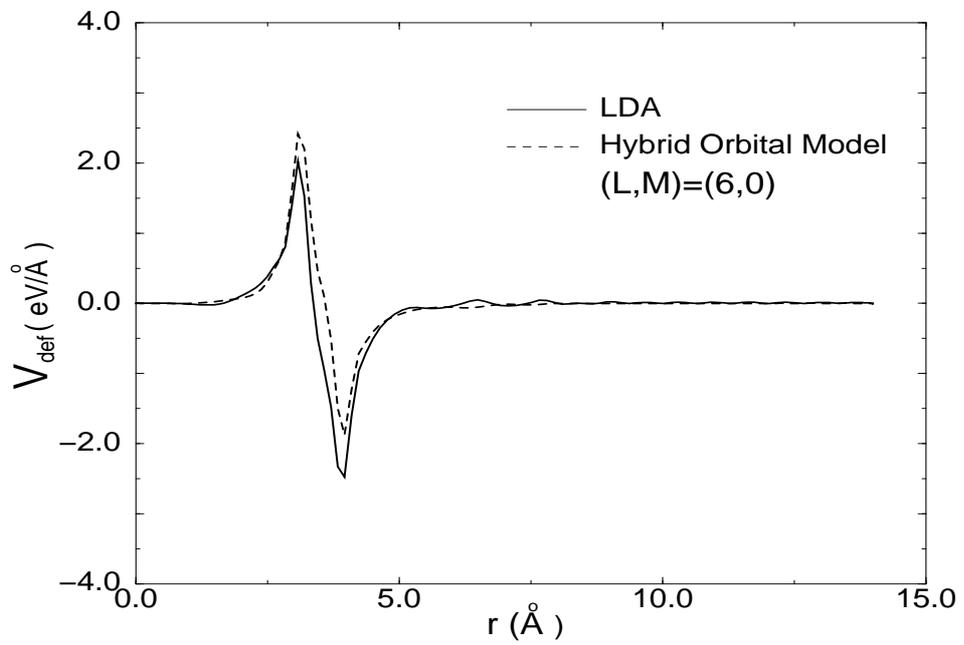
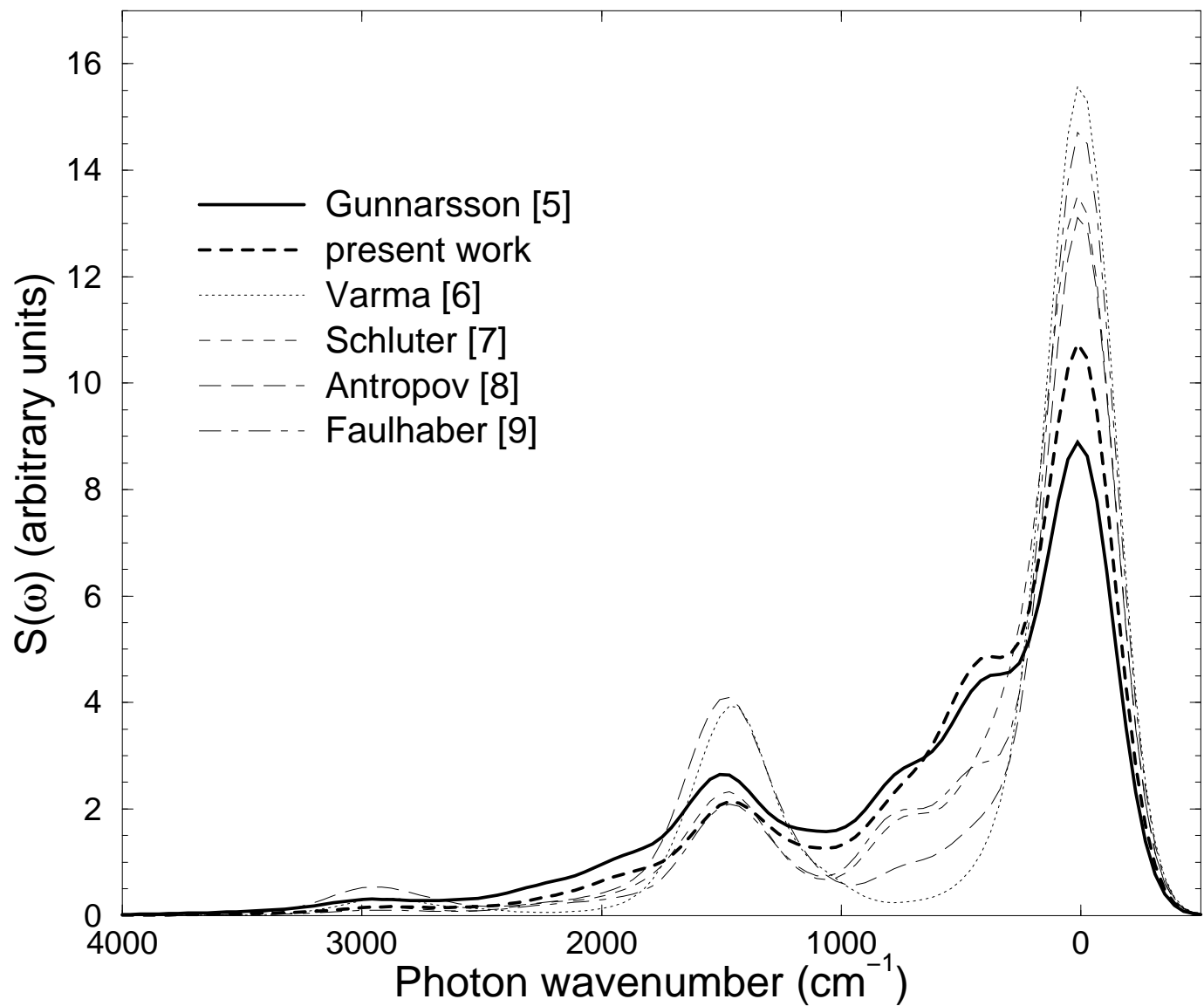
\end{document}